\documentclass[aps,prb,twocolumn,superscriptaddress]{revtex4}
\usepackage{ae}
\usepackage[T1]{fontenc}
\usepackage[ansinew]{inputenc}
\usepackage{amsmath}
\usepackage{amssymb}
\usepackage{graphicx}
\usepackage{color}

\usepackage[colorlinks]{hyperref}
\usepackage{epstopdf}

\def\sig{{\mbox{\boldmath{$\sigma$}}}}

\def\el{{\mbox{\boldmath{$\ell$}}}}

\def\sig{{\mbox{\boldmath{$\sigma$}}}}
\usepackage{soul,xcolor}

\def\sig{{\mbox{\boldmath{$\sigma$}}}}

\usepackage{soul,xcolor}
\setstcolor{red}

\def\sig{{\mbox{\boldmath{$\sigma$}}}}
\def\el{{\mbox{\boldmath{$\ell$}}}}
\def\mg{{\mbox{\boldmath{${\cal G}$}}}}
\def\NN{{\mbox{\boldmath{${\cal N}$}}}}

\makeatletter
\@ifundefined{textcolor}{}
{%
 \definecolor{BLACK}{gray}{0}
 \definecolor{WHITE}{gray}{1}
 \definecolor{RED}{rgb}{1,0,0}
 \definecolor{GREEN}{rgb}{0,1,0}
 \definecolor{BLUE}{rgb}{0,0,1}
 \definecolor{CYAN}{cmyk}{1,0,0,0}
 \definecolor{MAGENTA}{cmyk}{0,1,0,0}
 \definecolor{YELLOW}{cmyk}{0,0,1,0}
\definecolor{ORANGE}{rgb}{1,0,1} }

\@ifundefined{definecolor}
 {\@ifundefined{definecolor}
 {\@ifundefined{definecolor}
 {\@ifundefined{definecolor}
 {\@ifundefined{definecolor}
 {\usepackage{color}}{}
}{}
}{}
}{}
}{}

\usepackage{epsfig}

\def\sig{{\mbox{\boldmath{$\sigma$}}}}

\def\b0{{\bf{0}}}

\def\el{{\mbox{\boldmath{$\ell$}}}}

\begin{document}

\title{Effects of magnetic fields on the Datta-Das spin field-effect transistor }
\author{K. Sarkar}
\email{sarkark@post.bgu.ac.il}
\affiliation{Physics Department, Ben Gurion University, Beer Sheva 84105, Israel}
\affiliation{School of Physics and Astronomy, Tel Aviv University, Tel Aviv 69978, Israel}
\author{A. Aharony}
\email{aaharonyaa@gmail.com}
\affiliation{School of Physics and Astronomy, Tel Aviv University, Tel Aviv 69978, Israel}

\author{O. Entin-Wohlman}
\affiliation{School of Physics and Astronomy, Tel Aviv University, Tel Aviv 69978, Israel}

\author{M. Jonson}
\affiliation{Department of Physics, University of Gothenburg, SE-412
96 G{\" o}teborg, Sweden}

\author{R. I. Shekhter}
\affiliation{Department of Physics, University of Gothenburg, SE-412
96 G{\" o}teborg, Sweden}

\date{\today}

\begin{abstract}
A Datta-Das spin field-effect transistor is built of a heterostructure with a
Rashba spin-orbit interaction (SOI) at the interface (or quantum well)  separating two possibly magnetized reservoirs. The particle and spin
currents between the two reservoirs are driven by chemical potentials  that are (possibly) different for each spin direction.  These currents  are also tuned by varying  the strength of the SOI, which changes the amount of the rotation of the  spins of electrons crossing the heterostructure.
Here we  investigate the dependence of these currents on  additional Zeeman fields on the heterostructure and on variations of the reservoir magnetizations.  In contrast to the  particle current,  the spin currents are not necessarily conserved;  an additional spin polarization is injected into the reservoirs.
If a reservoir has a finite (equilibrium) magnetization, then we surprisingly find that the spin current into that reservoir can only have spins  which are parallel to the reservoir magnetization, independent of all the other fields. This spin current can be enhanced by increasing the magnetization of the other reservoir, and can also be tuned by the SOI and the various magnetic fields.
When only one reservoir is magnetized then the spin current into the other reservoir has  arbitrary tunable size and direction. In particular, this spin current changes  as the magnetization of the other reservoir is rotated. The optimal conditions for accumulating spin polarization on   an unpolarized reservoir are  to either apply a Zeeman field in addition to the SOI,   or to polarize the other reservoir.

 \end{abstract}


\maketitle


\section{Introduction}
\label{INTR}
Spin-polarized electrons can serve as mobile qubits that contain quantum information. A common way to tune the electron's spin polarization uses the spin-orbit interaction (SOI)\cite{winkler,manchon}. When an electron passes through a spin-orbit active material (e.g., semiconductor heterostructures \cite{Kohda}), its spin
 rotates around an effective magnetic field generated by the SOI. Both the direction of the rotation axis and the amount of rotation can be controlled in the case of the Rashba\cite{rashba} SOI  by appropriate tuning of gate voltages \cite{Nitta, Sato, Beukman,comDres}.

In a seminal paper\cite{datta}, Datta and Das proposed to use the SOI for the spin field-effect transistor (SFET):  The polarization of  electrons which come from one ferromagnetic reservoir (the source) rotates as they move through an SOI-active material (the link)  into another ferromagnetic reservoir (the drain). If the two ferromagnets have parallel magnetizations, if the electrons move ballistically and if the SOI is switched off, all the electrons enter the drain, in the `ON' state of the SFET. Switching on a gate voltage on the link, and tuning it so that the spins rotate by $180^o$ when they  reach the drain, the electrons are blocked from entering the drain, in the `OFF' state of the SFET. Although the literature contains many papers on possible realizations of the Datta-Das SFET \cite{DDhistory,recent,zutic1,flensberg},  most of these consider the Datta-Das SFET with fully and colinearly polarized conduction electrons in the reservoirs, and do not discuss the dependence of the particle and spin currents on the details of the (possibly partial) reservoir magnetizations or on additional magnetic fields. This analysis is presented below. In particular, we find an important difference between the case of two magnetized reservoirs, where  crucial restrictions on the spin polarization of the electrons appear, and the case where only one of the reservoirs is  magnetized, for which these restrictions do not exist in the other reservoir. We did not find such discussions in the existing literature.

\begin{figure}[htp]
\includegraphics[width=8cm]{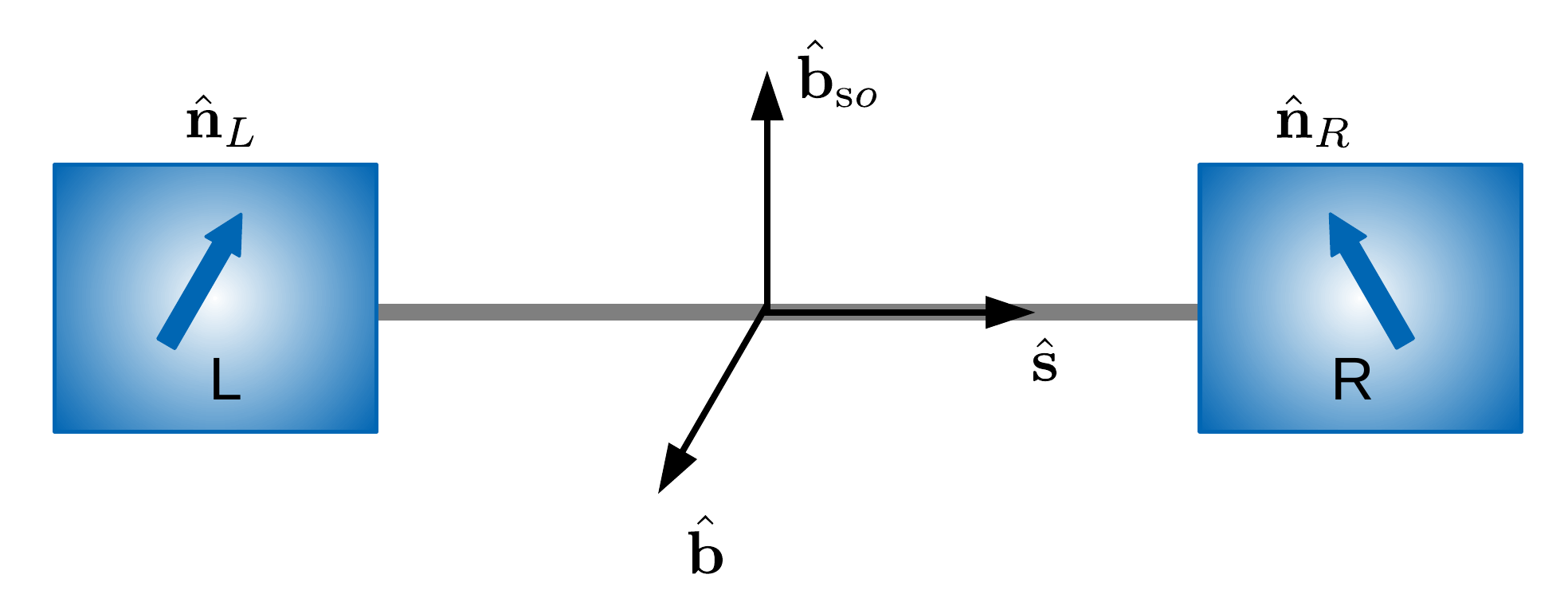}
\caption{(Color online.)
 The model: Two spin-polarized reservoirs, denoted by $L$ and $R$, have spin polarizations along the unit vectors $\hat{\bf n}^{}_L$ and $\hat{\bf n}^{}_R$ (indicated by the thick arrows). The reservoirs  are connected via a weak link,  represented by a one-dimensional wire (grey line),  which points along the unit vector $\hat{\bf s}$. The unit vector   $\hat{\bf b}_{\rm so}$ denotes the direction of an effective magnetic field induced by  the spin-orbit interaction on the wire.  An external magnetic field ${\bf B}=B\hat{\bf b}$ is also applied on the wire. }
\label{f1}
\end{figure}

In the simplest model of the Datta-Das device, the two reservoirs are connected by a one-dimensional wire (``weak link,"), see Fig. \ref{f1}. When the link is spin-orbit active, the single-channel,   two-terminal 2$\times$2  tunneling matrix (in spin space) through the link is unitary. Since time-reversal symmetry is obeyed, the transmission matrix is proportional to the unit matrix \cite{bardarson}, and  spin splitting cannot be achieved with SOI alone.
 In a recent paper \cite{Aharony_2018}, the time reversal symmetry was broken by a Zeeman energy gained from an external  magnetic field acting on the link. The   tunneling matrix through the link  is then non-unitary \cite{Aharony2_2019}, and spin splitting follows. For certain directions of
this field, both the charge and the spin conductances of the device were found to exhibit oscillations with the length of the weak link, even for unpolarized reservoirs.
Alternatively, time-reversal symmetry is broken when the leads are polarized \cite{Shekhter_2013,Shekhter_2014},   generalizing the Datta-Das ideal case. Some preliminary aspects of the reservoir polarizations were also reported in Ref. \onlinecite{Aharony1_2019}.
Reference \onlinecite{Shekhter_2014} had leads which were polarized only in the  longitudinal direction, with no magnetic field on the link. Reference \onlinecite{Aharony1_2019} had lead polarizations in general directions, but missed an important restriction on the polarizations generated by the spin-orbit interaction on the link. It also did not have a magnetic field on the link. These disadvantages, as well as a wrong sign in some of the results,  are removed in the present paper, which contains a comprehensive study of the effects of magnetic fields everywhere.

In this paper we study the charge and spin currents  through a spin-orbit active weak link, on which acts a Zeeman field, and consider various configurations of polarized reservoirs.  Most importantly, there is a crucial difference between an equilibrium reservoir magnetization, which is there even in the absence of spin currents, and a non-equilibrium magnetization, created by the driving forces of the currents (i.e., spin-dependent chemical potentials on the reservoirs, that can be generated by microwave irradiation). Somewhat surprisingly,  the spin current into a polarized lead is found to have spins which are polarized only along the existing magnetization.  Although this is a robust result of the calculation (correct to all orders in the tunneling energies\cite{keldysh}), its physical origin is not completely clear. Apparently,  spin states in the lead are quantized along the existing magnetization, and the incoming additional spins must adjust to that quantization axis.

After specifying the model Hamiltonian in Sec. \ref{model}, Sec. \ref{response} presents expressions for the particle and spin currents in the reservoirs, in terms of rate matrices which are derived in Sec. \ref{RatesM}. The  calculation is carried out  to second order in the tunneling matrix elements  in   the wide-band approximation \cite{Meir}.
Its most interesting outcome, i.e., the cancellation of the
 off-diagonal (in spin space) contributions, which were ignored in the earlier literature, can be   confirmed to all orders in the tunneling energies \cite{keldysh}. Using the explicit expression for the tunneling matrix \cite{Aharony2_2019}, and rotating the quantization axes of the reservoirs into general directions, Sec. \ref{res} presents explicit results for the particle and spin currents, to linear response in the spin-dependent chemical potentials, with the transport coefficients (generalized conductances) depending on the SOI, the Zeeman field acting on the link and on the reservoir polarizations. Details of the rotations and of the traces needed in the calculations are contained in Appendix \ref{traces}.  Section \ref{sum} summarizes our conclusions.

%
%


\section{Theoretical Model}
\label{model}

 Our system consists of two reservoirs connected by a Rashba-active weak link, as shown schematically in Fig.~\ref{f1}. The Hamiltonian of the entire system has contributions from the leads and from the tunneling of  electrons between the leads through the weak link,
\begin{align}
{\cal H}={\cal H}^{}_{\rm leads}+{\cal H}^{}_{\rm tun}\ .
\label{Htot}
\end{align}
In equilibrium, the electrons in each reservoir can be polarized in a ferromagnetic phase. The magnetization can also be generated by either  an external Zeeman field,  or by their band structure (e.g., half metals).  The creation (annihilation) operator of an electron with momentum ${\bf k}({\bf p})$ and spin $\sigma=+1,-1\equiv\uparrow,\downarrow$ in the left (right) lead is denoted  $c^{\dagger}_{{\bf k}({\bf p})\sigma}$ ($c^{}_{{\bf k}({\bf p})\sigma}$). The spin is quantized along the direction of the equilibrium spin polarization, i.e., along the unit vectors $\hat{\bf n}^{}_L$ and $\hat{\bf n}^{}_R$.  The  Hamiltonian of the left and right leads is
\begin{align}
{\cal H}_{\rm leads}=\sum_{{\bf k},\sigma}\epsilon^{}_{k\sigma}c^{\dagger}_{{\bf k}\sigma}c^{}_{{\bf k}\sigma}+ \sum_{{\bf p},\sigma}\epsilon^{}_{p\sigma}c^{\dagger}_{{\bf p}\sigma}c^{}_{{\bf p}\sigma}
\ ,
\label{Hl}
\end{align}
where $\epsilon^{}_{k(p)\sigma}$ is the (spin-dependent) energy of the electron. For instance, in  the presence of a Zeeman field   $\epsilon^{}_{k(p){\sigma}}=\epsilon^{}_{k(p)}- (g\mu^{}_B/2) H\sigma$, with the kinetic energy $\epsilon^{}_{k(p)}$.
Below we shall absorb the electronic magnetic moment $(-g\mu^{}_B/2)$ in the `magnetic field' $H$, which then has units of energy.

In addition to the (possible) equilibrium magnetization, spin polarization can be introduced through spin-dependent chemical potentials, e.g., by  irradiating the leads. Microwave radiation  can induce photon-assisted flips of the electrons' spins. The spin-polarization obtained this way is a non-equilibrium one. Assuming that the relaxation time of the electronic spin is much longer than the other relaxation processes,  the Fermi distribution functions can attain spin dependence \cite{Shekhter_2014, Aharony1_2019}  via   the chemical potentials,
\begin{align}
\mu^{}_{L(R)\sigma}=\mu^0_{L(R)}+\sigma U_{L(R)}\ .
\label{mu}
\end{align}
In principle, the magnetization due to $U^{}_{L(R)}$ need not be in the same direction as the equilibrium magnetization. For simplicity we assume here that it is quantized along the same direction $\hat{\bf n}^{}_{L(R)}$.

Next we describe the tunneling of electrons between the reservoirs  through the weak link. Denoting by  $[V^{}_{{\bf k}{\bf p}}]^{}_{\sigma\sigma'}$  the tunneling amplitude of the electron from the state with momentum ${\bf p}$ and spin $\sigma'$ in the right lead to the state with momentum ${\bf k}$ and spin $\sigma$ in the left lead,   the tunneling Hamiltonian is
\begin{align}
{\cal H}^{}_{\rm tun}=
\sum_{{\bf k},{\bf p},\sigma,\sigma'}([V^{}_{{\bf k}{\bf p}}]^{}_{\sigma\sigma'}c^{\dagger}_{{\bf k}\sigma}c^{}_{{\bf p}\sigma'}+{\rm H.c.})\ . 
\label{Htun}
\end{align}
As seen in Fig. \ref{f1}, the tunneling electron is subjected to an external  Zeeman field,
${\bf B}=B\hat{\bf b}$
(in energy units), and to the Rashba spin-orbit interaction, whose strength
is denoted $q_{\rm so}$ in momentum units (we use units in which $\hbar=1$). $q^{}_{\rm so}$ can be positive or negative, depending on the details of the SOI. The latter gives rise to an effective magnetic field
\begin{align}
{\bf B}^{}_{\rm so}({\bf q})=\frac{qq^{}_{\rm so}}{m^{\ast}_{}}\hat{\bf b}^{}_{\rm so}\ ,\ \ {\rm with} \ \ \ \hat{\bf b}^{}_{\rm so}=\hat{\bf n}\times\hat{\bf s}\ ,
\label{BQ}
\end{align}
whose direction
is normal to that of the electric field, $\hat{\bf n}$, generating the Rashba interaction, and the direction of the weak link
$\hat{\bf s}$, along which the electron, of mass $m^{\ast}$,  moves with momentum $q$.

The physics of the link is fully contained in the tunneling amplitude $V$. As explained in detail in App. B of Ref.~\onlinecite{Aharony2_2019}, this amplitude is obtained by evaluating the spin-dependent propagator between the ends of the tunneling region.
This propagator was calculated in Appendix B of Ref.~\onlinecite{Aharony2_2019}
for the  case
$\hat{\bf b}\perp\hat{\bf b}^{}_{\rm so}$, which is assumed here.
Denoting by $a$
 the extent of the localized wave function in the link \cite{raikh}, and adopting
the  plausible assumption that the  energy $[m^{\ast}a^{2}]^{-1}$  of an electron   there  is larger than the spin-orbit and the Zeeman energies,
\begin{align}
m^{\ast}Ba^{2}\ll 1\ ,\ \ q^{}_{\rm so}a \ll 1\ ,
\label{ineq}
\end{align}
one finds \cite{Aharony2_2019} that the tunneling amplitude can be written as $V^{}_{\bf kp}= J\widetilde{V}^{}_{LR}$, where $J$ (in energy units) is its value in the absence of the spin-orbit and Zeeman interactions, and
\begin{align}
\widetilde{V}^{}_{LR}=V^{}_0{\bf 1}+ iV^{}_{\rm so}(\hat{\bf b}_{\rm so}\cdot\sig)+V^{}_b(\hat{\bf b}\cdot\sig)
\label{Rot_General}
\end{align}
is a $2\times 2$ matrix, which contains the spin-dependence. Here, $\sig$ is the vector of the three Pauli matrices, ${\bf 1}$ is the $2\times 2$ unit matrix,
\begin{align}
V^{}_0=& \cos(q^{}_2 s)\ ,\ \  V^{}_{\rm so}=q^{}_{\rm so} [\sin(q^{}_2 s)/q^{}_2]\ ,\nonumber\\
V^{}_b=&m^{*}Ba[\sin(q^{}_2s)/q^{}_2]\ ,
\label{Vs}
\end{align}
$s$ is the length of the weak link, and
$q^{}_2\approx \sqrt{q_{\rm so}^2-(m^{*}Ba)^2}$.
In the absence of the Zeeman field
this becomes the Aharonov-Casher phase factor \cite{AC}, which is   a unitary matrix,
\begin{align}
\widetilde{V}^{}_{LR}\rightarrow e^{iq^{}_{\rm so} s\hat{\bf b}_{\rm so}\cdot\sig}\ .
\label{VV}
\end{align}
Since ${\bf b}^{}_{\rm so}=\hat{\bf n}\times\hat{\bf s}$ is normal to the direction $\hat{\bf n}$ of the electric field creating the spin-orbit interaction, and to the direction $\hat{\bf s}$ of the weak link, it follows that $\hat{\bf b}^{}_{{\rm so},LR}=-\hat{\bf b}^{}_{{\rm so},RL}$, and consequently
 $\widetilde{V}^{}_{RL}=\widetilde{V}^{\dagger}_{LR}$.


\section{ Particle and Spin Currents}
\label{response}

Both the particle and the spin currents in the left lead are determined by the rate
\begin{align}
&\sum_{\bf k}R^{L}_{\rm {\bf k}\sigma\sigma'}=\frac{d}{dt} \sum_{\bf k}\langle c^{\dagger}_{{\bf k}\sigma}c^{}_{{\bf k}\sigma'}\rangle
\ ,
\label{Heis}
\end{align}
with the angular brackets indicating quantum averaging. The particle current into the left lead is given by the diagonal terms (in spin space),
\begin{align}
I^{L}_{}=\sum_{{\bf k},\sigma}R^{L}_{{\bf k}\sigma\sigma}\ ,
\label{IL}
\end{align}
and the magnetization current (i.e., the spin current) is
\begin{align}
\dot{\bf M}^{L}_{}=\sum_{{\bf k},\sigma,\sigma'}R^{L}_{{\bf k}\sigma\sigma'}[\sig]^{}_{\sigma\sigma'}\ .
\label{MR}
\end{align}
The rate Eq. (\ref{Heis}) is conveniently separated into two contributions,
\begin{align}
&R^{L}_{\rm {\bf k}\sigma\sigma'}=R^{L,{\rm leads}}_{{\bf k}\sigma\sigma'}+R^{L,{\rm tun}}_{{\bf k}\sigma\sigma'}\ ,
\end{align}
where the first comes from the commutator with the leads' Hamiltonian,
\begin{align}
R^{L,{\rm leads}}_{{\bf k}\sigma\sigma'}& =
i\langle [{\cal H}^{}_{\rm leads},
c^{\dagger}_{{\bf k}\sigma}(t)c^{}_{{\bf k}\sigma'}(t)]\rangle \nonumber\\
&=
i(\epsilon^{}_{k\sigma}-\epsilon^{}_{k\sigma'})\langle c^{\dagger}_{{\bf k}\sigma}(t)c^{}_{{\bf k}\sigma'}(t)\rangle\ ,
\label{RLleads}
\end{align}
and the second from the commutator with the tunneling Hamiltonian,
\begin{align}
&R^{L,{\rm tun}}_{{\bf k}\sigma\sigma'} =
i\langle [{\cal H}^{}_{\rm tun},
c^{\dagger}_{{\bf k}\sigma}(t)c^{}_{{\bf k}\sigma'}(t)]\rangle\nonumber\\
&= i\sum_{{\bf p},\sigma^{}_{1}}
\Big (
[V^{\ast}_{{\bf k}{\bf p}}]^{}_{\sigma\sigma^{}_{1}}\langle c^{\dagger}_{{\bf p}\sigma^{}_{1}}(t)c^{}_{{\bf k}\sigma'}(t)-{\rm H.c.}\rangle\Big )\ .
\label{RkL}
\end{align}
 The time dependencies of the operators are  with respect to the full Hamiltonian Eq. (\ref{Htot}),  but   the rates themselves do not depend on time, as they pertain to a steady state of the system.
The corresponding expressions for the  rates belonging to the  right lead are found by replacing $L\rightarrow R$ and ${\bf k}\rightarrow{\bf p}$.
Current conservation is ensured when $
I^{L}_{}+I^{R}_{}=0$, which is indeed obeyed, since  $\widetilde{V}^{}_{RL}=\widetilde{V}^{\dagger}_{LR}$.
(see Sec. \ref{model}).


\section{Rate matrices}
\label{RatesM}

To second order in the tunneling matrix elements, the two contributions to the rate, Eqs. (\ref{RLleads}) and (\ref{RkL}),   are
\begin{widetext}
\begin{align}
R^{L,{\rm tun}}_{{\bf k}\sigma\sigma'}=i
\sum_{{\bf p},\sigma^{}_{2}}
[V^{\dagger}_{LR}]^{}_{\sigma^{}_{2}\sigma}
[V^{}_{LR}]^{}_{\sigma'\sigma^{}_{2}}\Big (\frac{f^{}_{L\sigma'}(\epsilon^{}_{k\sigma'})-f^{}_{R\sigma^{}_2}(\epsilon^{}_{p\sigma^{}_{2}})}{\epsilon^{}_{k\sigma'}-\epsilon^{}_{p\sigma^{}_{2}}-i\eta}+\frac{f^{}_{L\sigma}(\epsilon^{}_{k\sigma})
-f^{}_{R\sigma^{}_2}(\epsilon^{}_{p\sigma^{}_{2}})}{\epsilon^{}_{p\sigma^{}_{2}}-\epsilon^{}_{k\sigma}-i\eta}\Big )\ ,
\label{RLlead2}
\end{align}
and
\begin{align}
R^{L,{\rm lead}}_{{\bf k}\sigma\sigma'}=\frac{i(\epsilon^{}_{k\sigma}-\epsilon^{}_{k\sigma'})}{\epsilon^{}_{k\sigma}-\epsilon^{}_{k\sigma'}+i\eta}
\sum_{{\bf p},\sigma^{}_{2}}[V^{}_{LR}]^{}_{\sigma'\sigma^{}_{2}}[V^{\ast}_{LR}]^{}_{\sigma\sigma^{}_{2}}\Big (\frac{f^{}_{R\sigma^{}_2}(\epsilon^{}_{p\sigma^{}_{2}})-f^{}_{L\sigma'}(\epsilon^{}_{k\sigma'})}{\epsilon^{}_{k\sigma'}-\epsilon^{}_{p\sigma^{}_{2}}-i\eta}+\frac{f^{}_{R\sigma^{}_2}(\epsilon^{}_{p\sigma^{}_{2}})
-f^{}_{L\sigma}(\epsilon^{}_{k\sigma})}{\epsilon^{}_{p\sigma^{}_{2}}-\epsilon^{}_{k\sigma}-i\eta}\Big )\ ,
\label{RLl}
\end{align}
where $f_{L\sigma}$ and $f_{R\sigma}$ are the Fermi distributions in the leads [with the spin index indicating the spin dependence of the respective chemical potential, Eq. (\ref{mu})].
 The second contribution, Eq. (\ref{RLl}),  has only off-diagonal elements, $\sigma\ne\sigma'$, and it exists only when the left lead is polarized (at equilibrium), i.e., when  $\epsilon^{}_{k\sigma}\neq\epsilon^{}_{k\sigma'}$. In this case, the off-diagonal terms in the two contributions to   $R^{L}_{{\bf k}\sigma\sigma'}$  cancel each other, and this matrix becomes diagonal.  In fact, a Keldysh calculation within the wide-band approximation \cite{Meir} shows that this rather surprising cancellation happens to all orders in the tunneling.\cite{keldysh} This important result, which was not included in earlier papers,  implies that {\it the magnetization in a polarized lead can change only in the direction of its equilibrium polarization.}

Returning
 to Eq. (\ref{RLlead2}), we re-write it in the form
\begin{align}
R^{L,{\rm tun}}_{{\bf k}\sigma\sigma'}
=i\sum_{{\bf p},\sigma^{}_{2}}\int d\omega[V^{\ast}_{LR}]^{}_{\sigma\sigma^{}_{2}}&[V^{}_{LR}]^{}_{\sigma'\sigma^{}_{2}}\Big (\delta (\omega-\epsilon^{}_{p\sigma^{}_{2}})f^{}_{R\sigma^{}_{2}}(\omega)\Big [\frac{1}{\epsilon^{}_{k\sigma}-\omega+i\eta}-
\frac{1}{\epsilon^{}_{k\sigma'}-\omega-i\eta}
\Big ]\nonumber\\
&+\frac{f^{}_{L\sigma}(\omega)\delta(\omega-\epsilon^{}_{k\sigma})}{\epsilon^{}_{p\sigma^{}_{2}}-\omega-i\eta}-
\frac{f^{}_{L\sigma'}(\omega)\delta(\omega-\epsilon^{}_{k\sigma'})}{\epsilon^{}_{p\sigma^{}_{2}}-\omega+i\eta}\Big )
\ .
\label{RLtund}
\end{align}
Here, $f^{}_{L\sigma}(\omega)=[\exp[(\omega-\mu^{}_{L\sigma})/(k^{}_{\rm B}T)]+1]^{-1}$ is the Fermi distribution in the left lead [with an analogous definition for the distribution in the right reservoir, see Eq. (\ref{mu})].
\end{widetext}
The contributions from the principal parts of the integrals may be ignored (as can be seen by turning the sum over ${\bf p}$ into an integral, within the wide-band limit), we find
\begin{align}
\sum_{\bf k}R^{L,{\rm tun}}_{{\bf k}\sigma\sigma'}
&=\pi J^{2}\sum_{\sigma^{}_{2}}\int d\omega
[\widetilde{V}^{\ast}_{LR}]^{}_{\sigma\sigma^{}_{2}}[\widetilde{V}^{}_{LR}]^{}_{\sigma'\sigma^{}_{2}}
{\cal N}^{}_{R\sigma^{}_{2}}(\omega)
\nonumber\\
&\times\Big ({\cal N}^{}_{L\sigma}(\omega)[
f^{}_{R\sigma^{}_{2}}(\omega)-f^{}_{L\sigma}(\omega)]
\nonumber\\
&+{\cal N}^{}_{L\sigma'}(\omega)[
f^{}_{R\sigma^{}_{2}}(\omega)-f^{}_{L\sigma'}(\omega)]\Big )\ ,
\label{Not2}
\end{align}
where the density of states of the left lead is
\begin{align}
{\cal N}^{}_{L\sigma}(\omega)=\sum_{\bf k}\delta(\omega-\epsilon^{}_{k\sigma})\ ,
\end{align}
and similarly for is the density of states ${\cal N}_{R\sigma}(\omega)$ of the right lead.  In the following we treat these densities of states in the wide-band limit \cite{Meir}, replacing them by their values at the Fermi level.

The traces involved in the calculation of the currents, Eqs. (\ref{IL}) and (\ref{MR}),
are conveniently carried out in a matrix form, using the notations
\begin{align}
\NN^{}_{L(R)}={\cal N}^{0}_{L(R)}{\bf 1}+\Delta{\cal N}^{}_{L(R)}\sig\cdot{\bf n}^{}_{L(R)}\  ,
\end{align}
where
\begin{align}
{\cal N}^{0}_{L(R)}&=({\cal N}^{0}_{L(R)\uparrow}+{\cal N}^{0}_{L(R)\downarrow})/2\ , \nonumber\\
 \Delta{\cal N}^{}_{L(R)}&=({\cal N}^{0}_{L(R)\uparrow}-{\cal N}^{0}_{L(R)\downarrow})/2\ ,
\end{align}
and
\begin{align}
{\bf f}^{}_{L(R)}=\bar{f}^{}_{L(R)}{\bf 1}+\Delta f^{}_{L(R)}\sig\cdot\hat{\bf n}^{}_{L(R)}\ ,
\end{align}
where
\begin{align}
\bar{f}^{}_{L(R)}&=(f^{}_{L(R)\uparrow}+f^{}_{L(R)\downarrow})/2\ , \nonumber\\
    \Delta f^{}_{L(R)}&=(f^{}_{L(R)\uparrow}-f^{}_{L(R)\downarrow})/2\ .
\end{align}
Here we assumed that both the equilibrium and the non-equilibrium polarizations on the reservoirs (generated by $\Delta{\cal N}^{}_{L(R)}$ and by $\Delta f^{}_{L(R)}$, respectively) point in the same direction. It is easy to change that configuration via the rotations discussed in Sec. \ref{res}.
One then finds that the particle current  is
\begin{align}
I^L_{}=2\pi J^{2}&\int d\omega{\rm Tr}\Big \{\widetilde{V}^{}_{LR}\NN^{}_{R}{\bf f}^{}_{R}(\omega)
\widetilde{V}^{\dagger}_{LR}
\NN^{}_{L}
\nonumber\\
&-\widetilde{V}^{}_{LR}\NN^{}_{R}
\widetilde{V}^{\dagger}_{LR}
\NN^{}_{L}{\bf f}^{}_{L}(\omega)\Big \}\ ,
\label{IL2}
\end{align}
while the spin current in the left lead is
\begin{align}
&\dot{\bf M}^L_{}=\pi J^{2}\int d\omega{\rm Tr}\Big \{\Big(\widetilde{V}^{}_{LR}\NN^{}_{R}{\bf f}^{}_{R}(\omega)
\widetilde{V}^{\dagger}_{LR}
\NN^{}_{L}\nonumber\\
&+\NN^{}_{L}
\widetilde{V}^{}_{LR}\NN^{}_{R}{\bf f}^{}_{R}(\omega)
\widetilde{V}^{\dagger}_{LR}
-
\widetilde{V}^{}_{LR}\NN^{}_{R}
\widetilde{V}^{\dagger}_{LR}
\NN^{}_{L}{\bf f}^{}_{L}(\omega)\nonumber\\
&-\NN^{}_{L}{\bf f}^{}_{L}(\omega)\widetilde{V}^{}_{LR}\NN^{}_{R}
\widetilde{V}^{\dagger}_{LR}\Big )\sig\Big \}\ .
\label{ILs}
\end{align}
The corresponding currents in the right lead are found by changing $L\leftrightarrow R$, see Sec. \ref{model}.

In the linear-response regime, one expands the Fermi functions around their equilibrium value,
$f(\omega)=[e^{\beta(\omega-\mu)}+1]^{-1}$.
Assuming that the temperatures of the two leads are identical, we write
\begin{align}
f^{}_{L\sigma}(\omega)&\sim f(\omega)+(\mu^{}-\mu^{}_{L\sigma})\frac{\partial f(\omega)}{\partial\omega}\ ,
\end{align}
and use $\int d\omega\frac{\partial f(\omega)}{\partial\omega}=-1$, to obtain
\begin{align}
I^{L}_{}
&=G^{}_{0}(\mu^{0}_{R}-\mu^{0}_{L})-G^{}_{L,\parallel}U^{}_{L}+G^{}_{R,\parallel}U^{}_{R}\ ,
\label{IL1}
\end{align}
for the particle current,
with the transport coefficients
\begin{align}
G^{}_{0}&=2\pi J^{2}_{} {\rm Tr}\{\widetilde{V}^{\dagger}_{LR}\NN^{}_{L}\widetilde{V}^{}_{LR}\NN^{}_{R}\}\ ,\nonumber\\
G^{}_{L,\parallel}&= 2\pi J^{2}_{}{\rm Tr}\{\widetilde{V}^{\dagger}_{LR}\NN^{}_{L}\sig\cdot\hat{\bf n}^{}_L\widetilde{V}^{}_{LR}\NN^{}_{R}\}\ ,\nonumber\\
G^{}_{R,\parallel}&=2\pi J^{2}_{} {\rm Tr}\{\widetilde{V}^{}_{LR}\NN^{}_{R}\sig\cdot\hat{\bf n}^{}_R\widetilde{V}^{\dagger}_{LR}\NN^{}_{L}\}\ .
\label{Gz}
\end{align}
Interchanging $L$ and $R$ also implies interchanging $\widetilde{V}_{LR}$ and $\widetilde{V}^\dagger_{LR}$, hence $I^{L}_{}=-I^{R}_{}\equiv I$; the total number of particles is conserved.

The linear-response expression for the spin current  in the left lead is
\begin{align}
\dot{\bf M}^{L}_{}
&=\pi J^{2}{\rm Tr}\{\widetilde{V}^{}_{LR}\NN^{}_{R}\widetilde{V}^{\dagger}_{LR}(\NN^{}_L\sig+\sig\NN^{}_L)(\mu_R^0-\mu_L^0)\nonumber\\
&-
\widetilde{V}^{}_{LR}\NN^{}_{R}\widetilde{V}^{\dagger}_{LR}(\NN^{}_L\sig\cdot\hat{\bf n}^{}_L\sig+\sig\NN^{}_L\sig\cdot\hat{\bf n}^{}_L)U^{}_L\nonumber\\
&+\widetilde{V}^{}_{LR}\NN^{}_{R}\sig\cdot\hat{\bf n}^{}_R\widetilde{V}^{\dagger}_{LR}(\NN^{}_L\sig+\sig\NN^{}_L)U^{}_R\}\ .
\label{ILs1}
\end{align}
As discussed above,  when the lead $L$ is polarized then only the magnetization parallel to the quantization axis $\hat{\bf n}^{}_L$  survives. One then finds
\begin{align}
\dot{M}^L_\parallel&=G^{}_{L,\parallel}(\mu^0_R-\mu^0_L)-G^{}_0U^{}_L+G^{}_{\times,\parallel}U^{}_R\ ,
\label{Mpar}
\end{align}
where
\begin{align}
G^{}_{\times,\parallel}&=2\pi J^{2}{\rm Tr}\{\widetilde{V}^{\dagger}_{LR}\NN^{}_{L}\sig\cdot{\bf n}^{}_L\widetilde{V}^{}_{LR}\NN^{}_{R}\sig\cdot\hat{\bf n}^{}_R\}\ ,
\label{xpara}
\end{align}
 and the other transport coefficients are given in Eqs. (\ref{Gz}).
Equations (\ref{IL1}) and (\ref{Mpar}) generalize Eq. (23) of Ref. \onlinecite{Shekhter_2014}, by the addition of the magnetic field acting on the weak link.

 When the lead $L$ is unpolarized, there is no meaning to the choice of the  quantization axis along $\hat{\bf n}^{}_L$. Instead, one introduces the direction of the magnetization generated by $U^{}_L$,
denoted (for simplicity) by the same $\hat{\bf n}^{}_L$ (see Fig. \ref{f1}), and then
\begin{align}
\dot{\bf M}^L&={\bf G}^{}_{L}(\mu^0_R-\mu^0_L)-G^{}_0U^{}_L\hat{\bf n}^{}_L+{\bf G}^{}_{\times}U^{}_R\ .
\label{MML}
\end{align}
The new transport coefficients ${\bf G}^{}_{L}$ and ${\bf G}^{}_{\times}$ can be read from Eq. (\ref{ILs1}). They are elaborated upon in Sec. \ref{res}.


\section{Results}
\label{res}

The detailed calculations of the particle and spin currents in terms of the various chemical potentials and spin-dependent densities of state is presented in App. \ref{traces}.
Several specific configurations of the polarization axes can be considered.  In the examples described below, we choose the link to lie along the ${\bf x}-$axis ($\hat{\bf s}=\hat{\bf x}$), the Zeeman field acting on the weak link to be directed along the ${\bf z}-$axis ($\hat{\bf b}=\hat{\bf z}$),  and the effective magnetic field induced by the spin-orbit coupling  along the ${\bf y}-$axis ($\hat{\bf b}^{}_{\rm so}=\hat{\bf y}$), so that   $\hat{\bf b}^{}_{\rm so}\times\hat{\bf b}=\hat{\bf s}$. Recall that we use units in which $\hbar=1$, and consider particle (and not charge) currents.


\subsection{\bf Unpolarized reservoirs}
\label{subA}

\begin{widetext}

\begin{figure}[htp]
\includegraphics[width=5.5cm]{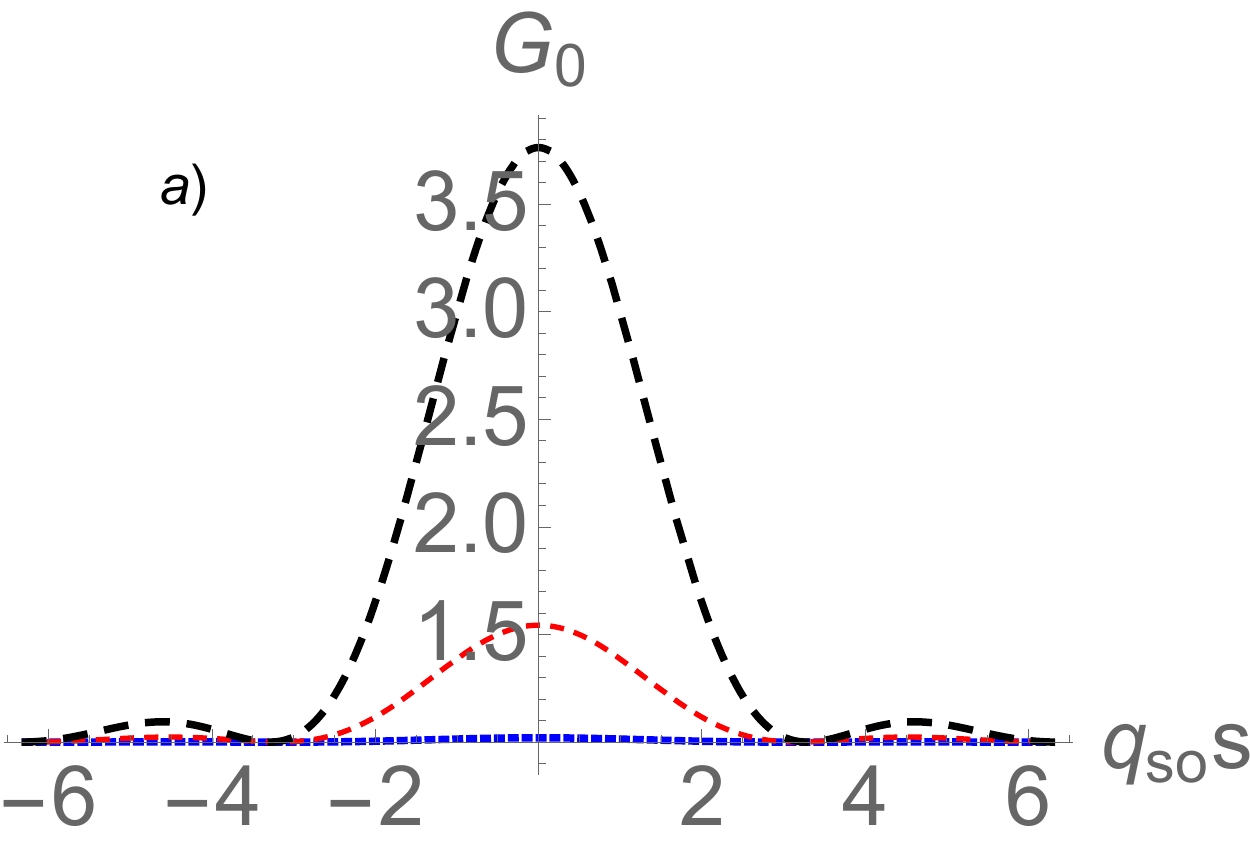}\ \ \ \
\includegraphics[width=5.5cm]{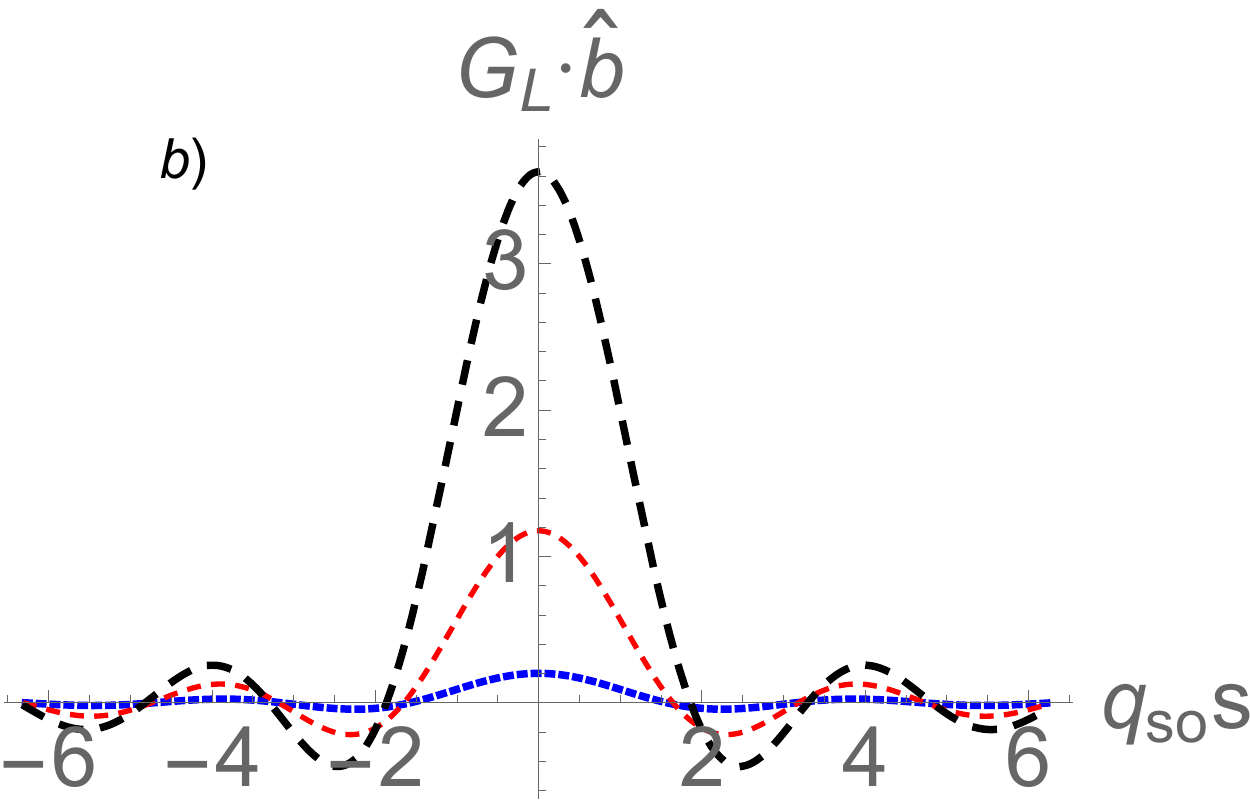}\ \ \ \
\includegraphics[width=5.5cm]{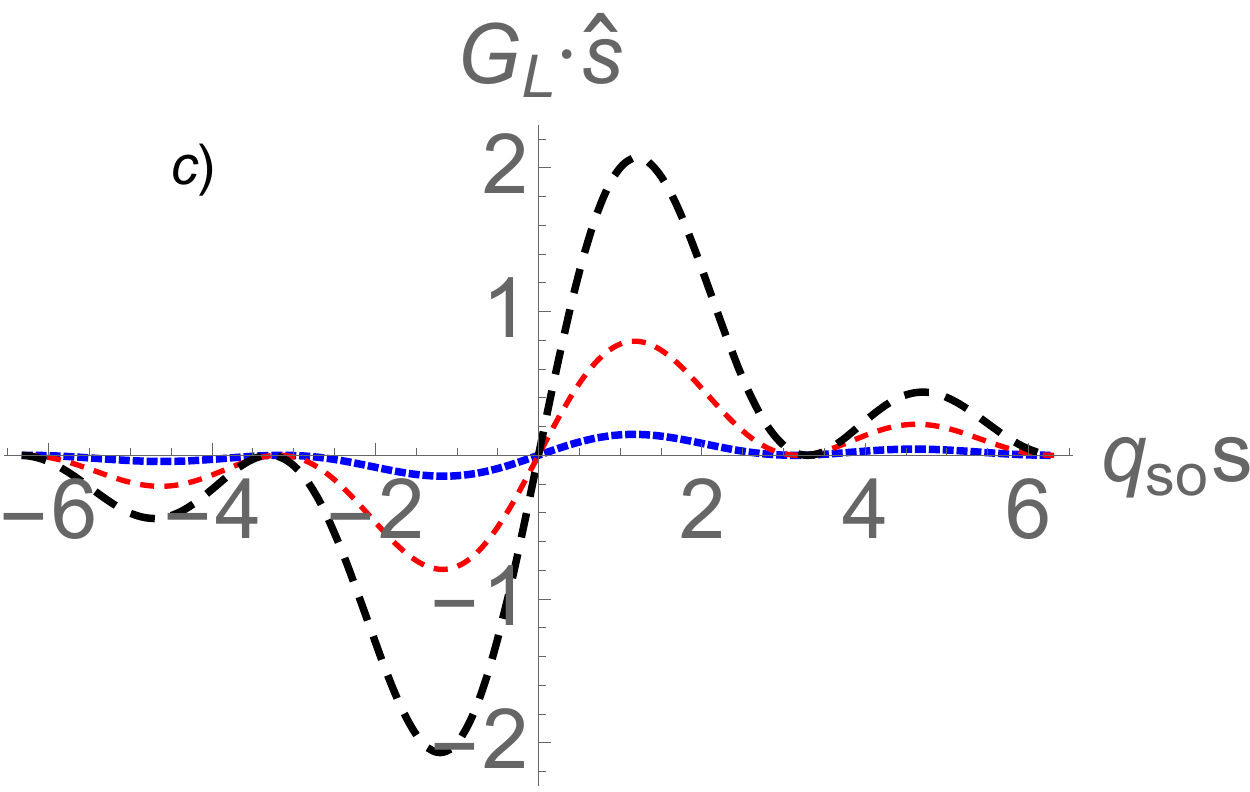}
\caption{(Color online.)
Conductance coefficients (in units of $\gamma=4\pi J^2{\cal N}^0_L{\cal N}^0_R$) versus the strength $q^{}_{\rm so}$ of the spin-orbit coupling (in units of $1/s)$, or the length of the link $s$ (in units of $1/q^{}_{\rm so}$) for unpolarized leads. The three lines are for $m^\ast aBs=0.1,~0.5,~1$ (blue, red, black, with increasing dashes). (a) $G^{}_0(B)$, Eq. (\ref{G0unp}),
(b) ${\bf G}^{}_L\cdot\hat{\bf b}$ and  (c) ${\bf G}^{}_L\cdot\hat{\bf s}$, Eq. (\ref{55}). }
\label{unp}
\end{figure}
\end{widetext}

In the simplest configuration, both leads in the decoupled junction are not polarized,  the densities of states are independent of the spin index, i.e., $\Delta{\cal N}^{}_L=\Delta{\cal N}^{}_R=0$, and the (dimensionless) conductance $G_{0}$ becomes [see Eqs. (\ref{Rot_General}) and (\ref{Vs})]
\begin{align}
G^{}_0&=\gamma\mathcal{G}^{}_0=\gamma(V_0^2+V_{\rm so}^2+V_b^2)\ ,
\label{G0unp}
\end{align}
where
\begin{align}
\gamma=4\pi J^2{\cal N}^0_L{\cal N}^0_R
\end{align}
is the  conductance (or transparency) of the link without the spin-orbit and the Zeeman interactions, i.e., when $V^{}_{\rm so}=V^{}_b=0$ and $V^{}_0=1$.
The conductance (\ref{G0unp}) 
does not depend on the directions of these fields.  It monotonically increases with $B$ and oscillates with $q^{}_{\rm so}s$, see Fig. \ref{unp}(a). Since the reservoirs are unpolarized,
the vectors $\hat{\bf n}^{}_L$ and $\hat{\bf n}^{}_R$ are not defined (unless $U^{}_L$ and/or $U^{}_R$ are non-zero, see below). Instead, Eqs. (\ref{Gz}) and (\ref{C5}) show that
\begin{align}
{\bf G}^{}_L&=(\gamma/2){\rm Tr}\{\widetilde{V}^\dagger_{LR}\sig \widetilde{V}^{}_{LR}\}=\gamma(\mg^{}_1-\mg^{}_2)\nonumber\\
&=2V_b\gamma(V^{}_0\hat{\bf b}-V^{}_{\rm so}\hat{\bf b}^{}_{\rm so}\times\hat{\bf b})\ .
\label{55}
\end{align}
With our choice of the directions, the second term points along the link. The magnetization generated in the left lead by a bias voltage $\mu^0_R-\mu^0_L$ (note: we use $e=1$) then has a component along $\hat{\bf b}$, of magnitude $2\gamma V^{}_0V^{}_b$, which is odd in ${\bf B}$ and even in $q^{}_{\rm so}$
 [Fig. \ref{unp}(b)], and a component along $\hat{\bf s}$, of magnitude $2\gamma V^{}_{\rm so}V^{}_b$, which is odd in ${\bf B}$ and  in $q^{}_{\rm so}$ [Fig. \ref{unp}(c)]. Both components vanish in the absence of the Zeeman field (as expected \cite{bardarson} when time-reversal symmetry is restored), and grow monotonically with $|{\bf B}|$.  Varying  the electric field which determines the strength of the spin-orbit coupling at a fixed value of $B$, rotates   this magnetization in the $\hat{\bf b}-\hat{\bf s}$ plane (see Fig. \ref{f1}), which is perpendicular to the SOI vector $\hat{\bf b}^{}_{\rm so}$. From Eqs. (\ref{GGGG}) and (\ref{C4}) one can also see that the corresponding coefficient for the right reservoir,   ${\bf G}^{}_R=\gamma(\mg^{}_1+\mg^{}_2)$,  is given by the same components, except that the component along the link changes sign.

The above results represent the currents for $U^{}_L=U^{}_R=0$ [see Eq. (\ref{mu})]. However, spin polarization in the leads can also be induced when the chemical potentials   assigned to them are  spin-dependent. These polarizations are associated with a direction of the magnetization, $\hat{\bf n}^{}_L$ and $\hat{\bf n}^{}_R$ for the left and right lead, respectively. The contributions to the particle current are then given by Eq. (\ref{IL1}), with
$G^{}_{L,\parallel}={\bf G}^{}_L\cdot\hat{\bf n}^{}_L,~
G^{}_{R,\parallel}={\bf G}^{}_R\cdot\hat{\bf n}^{}_R$.
The magnetization current is given by Eq. (\ref{MML}),
with ${\bf G}^{}_\times=\gamma \mg^{\hat{\bf n}^{}_R}_3$.  This vector modifies the rotation of the magnetization of the electrons as they move from the right lead to the left lead through the weak link, which contains both the SOI and the Zeeman fields. In particular, $\mg^{\hat{\bf n}^{}_R}_3=(V_0^2-V_{\rm so}^2+V_b^2)\hat{\bf b}-2V^{}_0V^{}_{\rm so}\hat{\bf s},~(V_0^2-V_b^2-V_{\rm so}^2)\hat{\bf s}+2V^{}_0V^{}_{\rm so}\hat{\bf b},~(V_0^2-V_b^2+V_{\rm so}^2)\hat{\bf b}^{}_{\rm so},$ for $\hat{\bf n}^{}_R=\hat{\bf b},~\hat{\bf s},~\hat{\bf b}^{}_{\rm so}$, and all these components oscillate with $q^{}_2 s$ (which decreases as $|{\bf B}|$ increases). In the first two cases this vector contributes to the rotation of the spin current in the $\hat{\bf b}-\hat{\bf s}$ plane, in addition to the results shown in  Figs. \ref{unp}(b) and (c). In the third case, this vector generates a spin component perpendicular to that plane, i.e., along its original direction along $\hat{\bf n}^{}_R$.

\subsection{\bf Two polarized reservoirs}
\label{2p}

When both leads are polarized, there are no `transverse' components of the spin currents. In equilibrium, the leads are polarized along $\hat{\bf n}^{}_L$ and $\hat{\bf n}^{}_R$, with magnetizations which are proportional to $p^{}_{L}\equiv \Delta{\cal N}^{}_{L}/{\cal N}^0_{L}$
and
$p^{}_{R}\equiv \Delta{\cal N}^{}_{R}/{\cal N}^0_{R}$, respectively.  In the absence of the  Zeeman field on the link, we find that  $\mg^{}_1=\mg^{}_2=0$ and $\mathcal{G}^{}_4=
(V_0^2-V_{\rm so}^2)
\hat{\bf n}^{}_L\cdot\hat{\bf n}^{}_R+2V_{\rm so}^2(\hat{\bf b}^{}_{\bf so}\cdot\hat{\bf n}^{}_R)(\hat{\bf b}^{}_{\bf so}\cdot\hat{\bf n}^{}_L)
-2V^{}_0V^{}_{\rm so}[\hat{\bf n}^{}_R\times\hat{\bf n}^{}_L]\cdot\hat{\bf b}^{}_{\rm so}$, and consequently
\begin{align}
&G^{}_0=\gamma\big(\mathcal{G}_0+p^{}_Lp^{}_R\mathcal{G}_4\big)\ ,
\nonumber\\
&G^{}_{L,\parallel}=\gamma\big(p^{}_R\mathcal{G}_4+p^{}_L\mathcal{G}_0\big)\ ,\nonumber\\
&G^{}_{R,\parallel}=\gamma\big(p^{}_L\mathcal{G}_4+p^{}_R\mathcal{G}_0\big)\ ,\nonumber\\
&G^{}_{\times,\parallel}=\gamma\big(p^{}_Lp^{}_R\mathcal{G}_0+\mathcal{G}_4\big)\ .
\end{align}
If the SOI interaction also vanishes, we have $\mathcal{G}_0=1$  and $\mathcal{G}_4=\hat{\bf n}^{}_L\cdot\hat{\bf n}^{}_R$, and then
\begin{align}
&G^{}_0=\gamma\big(1+p^{}_Lp^{}_R\hat{\bf n}^{}_L\cdot\hat{\bf n}^{}_R\big)\ ,
\nonumber\\
&G^{}_{L,\parallel}=\gamma\big(p^{}_R\hat{\bf n}^{}_L\cdot\hat{\bf n}^{}_R+p^{}_L\big)\ ,\nonumber\\
&G^{}_{R,\parallel}=\gamma\big(p^{}_L\hat{\bf n}^{}_L\cdot\hat{\bf n}^{}_R+p^{}_R\big)\ ,\nonumber\\
&G^{}_{\times,\parallel}=\gamma\big(p^{}_Lp^{}_R+\hat{\bf n}^{}_L\cdot\hat{\bf n}^{}_R\big)\ .
\end{align}
Note that when the leads are  unpolarized only the `usual' transparency of the link, $G^{}_0=\gamma$,  survives, and there are no spin currents.
Choosing $p^{}_L,~p^{}_R>0$, the conductance and the spin conductance are maximal when $\hat{\bf n}^{}_L=\hat{\bf n}^{}_R$, and minimal when $\hat{\bf n}^{}_L=-\hat{\bf n}^{}_R$. In particular, in the latter case the current is completely blocked when the leads are fully polarized, $p^{}_L=p^{}_R=1$, as expected by the theory of Datta and Das \cite{datta}.

A configuration where the leads are   fully polarized and $p^{}_L=p^{}_R=1$, is realized for instance when they are made of half metals.
In this case,
\begin{align}
G^{}_0=G^{}_{L,\parallel}&=G^{}_{R,\parallel}=G^{}_{\times,\parallel}=\gamma\big[(\mg^{}_1-\mg^{}_2)\cdot\hat{\bf n}^{}_L\nonumber\\
&+(\mg^{}_1+\mg^{}_2)\cdot\hat{\bf n}^{}_R+\mathcal{G}^{}_0+\mathcal{G}^{}_4\big]\ .
\label{polG}
\end{align}
The equality of all these coefficients simply means that
$I^L_{}=G^{}_0(\mu^{}_{R\uparrow}-\mu^{}_{L\uparrow})$, where $\mu^{}_{L(R)\uparrow}=\mu^0_{L(R)}+U^{}_{L(R)}$.
When $\hat{\bf n}^{}_L$ and $\hat{\bf n}^{}_R$ are both parallel to $\hat{\bf b}$, this reduces to
\begin{align}
G^{}_0=G^{}_{L,\parallel}=G^{}_{R,\parallel}=G^{}_{\times,\parallel}=2\gamma\big(V^{}_0+V^{}_b\big)^2\ .
\end{align}
This expression applies when $\hat{\bf n}^{}_L=\hat{\bf n}^{}_R=\hat{\bf b}$. If we change the sign of $\hat{\bf b}$, this expression becomes $\gamma\big(V^{}_0-V^{}_b\big)^2$. In both cases,  the Zeeman field can be used to increase or decrease both the charge and the spin conductances.

For the other possibility,     $\hat{\bf n}^{}_L=-\hat{\bf n}^{}_R=\hat{\bf b}$, we obtain
\begin{align}
G^{}_0=G^{}_{L,\parallel}=G^{}_{R,\parallel}=G^{}_{\times,\parallel}=2\gamma V_{\rm so}^2\ .
\end{align}
Indeed, in this case the SOI opens the Datta-Das blocking, and the conductances oscillate strongly with $q^{}_{\rm so}$. Without the SOI there is no (particle or spin) current between the reservoirs. Interestingly, this result depends on the Zeeman field only via $q^{}_2$, which determines the period of the oscillations.

When $\hat{\bf n}^{}_L=\hat{\bf n}^{}_R=\hat{\bf s}=\hat{\bf b}^{}_{\rm so}\times \hat{\bf b}$, we have
\begin{align}
G^{}_0=G^{}_{L,\parallel}=G^{}_{R,\parallel}=G^{}_{\times,\parallel}=2\gamma V_0^2\ .
\end{align}
 Clearly, rotating the two reservoir magnetizations allows the measurement of all three coefficients in Eq. (\ref{Vs}), $V^{}_0,~V^{}_b$ and $V^{}_{\rm so}$.



\subsection{\bf A polarized lead coupled to an unpolarized one}
\label{hp}

\begin{figure}[htp]
\includegraphics[width=7cm]{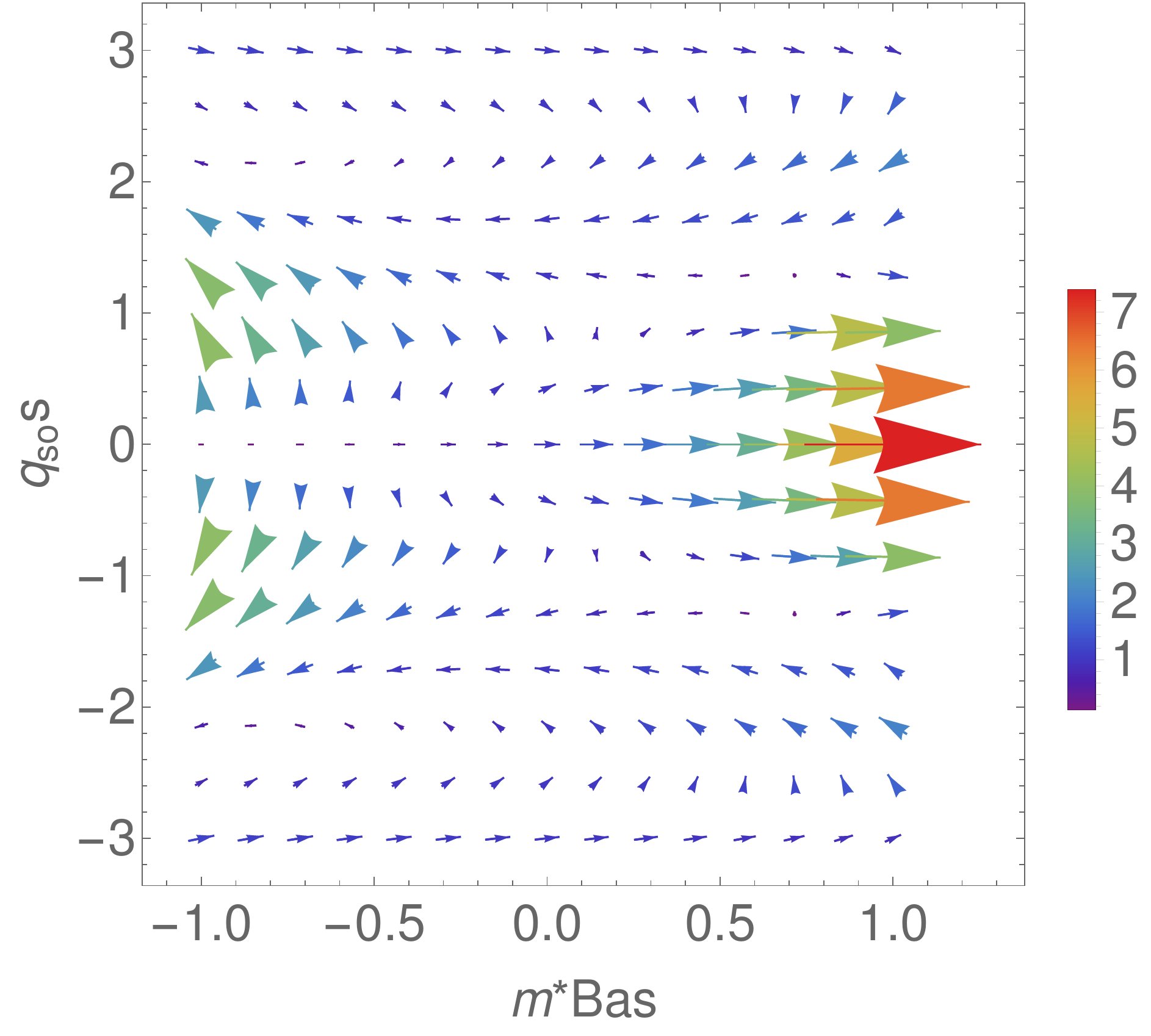}
\caption{(Color online.)
The magnetization conductance ${\bf G}^{}_L$ (in units of $\gamma$) injected into the left un-polarized lead ($p^{}_L=0$) due to a full polarization of the right lead ($p^{}_R=1$), for different values of the SOI and the Zeeman energy on the link, with $\hat{\bf n}^{}_R=\hat{\bf b}$. The arrows are all in the $\hat{\bf b}-\hat{\bf s}$ plane. }
\label{onelead}
\end{figure}

Assuming that only the right lead is polarized, the transport coefficients for the charge and spin currents in that lead are  given by Eqs. (\ref{GGGG}), alas with $\Delta{\cal N}^{}_L=0$. However, the vector $\hat{\bf n}^{}_L$ loses its meaning, and there is no distinction between the `longitudinal' and `transverse' spin components in the transport coefficients pertaining to the  unpolarized lead $L$. Instead, combining Eqs. (\ref{GGGG}) and (\ref{Mx}) or a direct derivation based on Eq. (\ref{Gz}) give
\begin{align}
\dot{\bf M}^L_{}={\bf G}^{}_L(\mu^0_R-\mu^0_L)+{\bf G}^{}_\times U^{}_R\ ,
\label{54}
\end{align}
with
\begin{align}
G^{}_0&=\gamma[\mathcal{G}^{}_0+p^{}_R(\mg^{}_1+\mg^{}_2)\cdot\hat{\bf n}^{}_R]\ ,\nonumber\\
{\bf G}^{}_L&=\gamma\big(\mg^{}_1-\mg^{}_2+p^{}_R\mg^{\hat{\bf n}^{}_R}_3\big)\ ,\nonumber\\
{\bf G}^{}_\times&=\gamma[\mg^{\hat{\bf n}^{}_R}_3+p^{}_R(\mg^{}_1-\mg^{}_2)]\ .
\end{align}
The first term in the expression for ${\bf G}^{}_L$ coincides with Eq. (\ref{55}) for the unpolarized leads, and the second term represents the additional contribution from the polarization of the right lead. When the SOI and the Zeeman field on the link vanish then $\mg^{\hat{\bf n}^{}_R}_3=\hat{\bf n}^{}_R$, and the result ${\bf G}^{}_L=\gamma p^{}_R\hat{\bf n}^{}_R$ reflects the polarization on the lead $R$, as expected.
The transport coefficients for the unpolarized case were already presented in Fig. \ref{unp},  and examples of $\mg^{\hat{\bf n}^{}_R}_3$, for several directions of $\hat{\bf n}^{}_R$,  were discussed at the end of Sec. \ref{subA}.
Interestingly, the effects of $p^{}_R$ and of $U^{}_R$ on the spin current in lead $L$ are similar: both generate an additional spin current along $\mg^{\hat{\bf n}^{}_R}_3$. The values of ${\bf G}^{}_L$, for $p^{}_R=1$ and $\hat{\bf n}^{}_R=\hat{\bf b}$, are shown in Fig. \ref{onelead}.

\section{Summary and Conclusions}
\label{sum}

In this paper we have  systematically investigated the effects of a Zeeman field on the particle and spin currents through a spin-orbit active weak link connecting two reservoirs, in the possible presence of equilibrium magnetizations in the reservoirs.
Even when the reservoirs are not polarized, the magnetic field acting in  the link increases the particle conductance through the link, and generates tunable spin currents flowing into the reservoirs. Adding a non-equilibrium magnetization in one of the reservoirs (by spin-dependent chemical potentials) allows further tuning of the magnitude and direction of the magnetization injected into the other (unpolarized) reservoir.

The original Datta-Das spin field-effect transistor \cite{datta} assumed that the reservoirs are fully polarized, and this can be achieved, e.g. with half metals. Without the magnetic field acting in the link,  the SOI reproduces the Datta-Das predictions, e.g. lifting the spin blocking which arises when the reservoirs are polarized in opposite directions. Here we generalized their model, and allowed for  arbitrary directions of the magnetizations of the reservoirs. Our most surprising result (which holds to all orders in the perturbation expansion) is that  spins injected into a polarized reservoir can only modify this polarization, without any `transverse' components. For this reason, we recommend polarizing only one reservoir. In this case we find that spin injection into the other reservoir is completely tunable, and we give explicit simple expressions for this spin current. Both an equilibrium polarization and  polarization due to spin-dependent chemical potential on the right reservoir  generate an additional left reservoir magnetization along $\mg^{\hat{\bf n}^{}_R}_3$ [defined in Eq. (\ref{GG3})], which adds more tuning possibilities of the latter magnetization.

Without any time-reversal-breaking fields there is no injection of spin polarization into the reservoirs. If the aim of the spintronic device is to produce such a polarization, then one must add a magnetic field and/or  magnetically polarize some parts of the system. Our results show that this is best achieved if one adds a Zeeman field that acts on the link, without polarizing the reservoirs, or by polarizing one of the reservoirs (preferably using a half metal) and measuring the spin accumulation in the other reservoir. In the former case, the spin polarizations in the two reservoirs differ from each other, and can be tuned separately.

We do not have a full `hand-waving' explanation for the vanishing of the `transverse' spin components in the polarized reservoirs. Apparently, the equilibrium polarization sets a quantization axis for the spins along that polarization, and the incoming spins are projected onto that axis, while their transverse components average to zero.

The strength of the spin-orbit interaction appears in our results via the dimensionless product $q^{}_{\rm so}s$ [Eq. (\ref{Vs}) and following text]. Are there materials for which this product can be significant?
A Shubnikov-de Haas experiment\cite{69} on an Al$_{0.25}$In$_{0.75}$As barrier layer
gave a value for the Rashba coefficient (in different units)
of $\alpha = 3 \times 10^{-11}$ eV/m. With the effective mass $m^\ast =0.023$ m$_0$, this gives $q^{}_{\rm so} = m^\ast\alpha/\hbar^2 = 9 \times 10^6$ m$^{-1}$. Weak antilocalization measurements in a quaternary InGaAsP/InGaAs
heterointerface\cite{70} yielded $\alpha = 10.4 \times 10^{-12}$ eV/m. With an
effective mass $m^\ast = 0.0408$ m$_0$, this gives $q^{}_{\rm so}= 5.55 \times 10^6$ m$^{-1}$. Thus, $s = 300$ nm would imply 
$q^{}_{\rm so}s \approx 1.6 - 2.7$,
which is quite large.


\acknowledgments

This research was  partially supported by the Israel Science Foundation (ISF), by the infrastructure program of Israel Ministry of Science and Technology under contract 3-11173, and by the Pazy Foundation. We acknowledge the hospitality of the PCS at IBS, Daejeon, Korea, where part of this work was  supported by
IBS funding number
(IBS-R024-D1), and Zhejiang University, Hangzhou, China.


\appendix


\section{Traces}
\label{traces}

The transport coefficients in Eq. (\ref{Gz}) are
\begin{widetext}
\begin{align}
G^{}_0/(2\pi J^2)={\rm Tr}\{\widetilde{V}^{\dagger}_{LR}({\cal N}^{0}_{L}{\bf 1}+\Delta{\cal N}^{}_L\hat{\bf n}^{}_L\cdot\sig)\widetilde{V}^{}_{LR}({\cal N}^{0}_{R}{\bf 1}+\Delta{\cal N}^{}_R\hat{\bf n}^{}_R\cdot\sig)\}\ ,\nonumber\\
G^{}_{L,\parallel}/(2\pi J^2)={\rm Tr}\{\widetilde{V}^{\dagger}_{LR}({\cal N}^{0}_{L}\hat{\bf n}^{}_L\cdot\sig+\Delta{\cal N}^{}_L{\bf 1})\widetilde{V}^{}_{LR}({\cal N}^{0}_{R}{\bf 1}+\Delta{\cal N}^{}_R\hat{\bf n}^{}_R\cdot\sig)\}\ ,\nonumber\\
G^{}_{R,\parallel}/(2\pi J^2)={\rm Tr}\{\widetilde{V}^{}_{LR}({\cal N}^{0}_{R}\hat{\bf n}^{}_R\cdot\sig+\Delta{\cal N}^{}_R{\bf 1})\widetilde{V}^{\dagger}_{LR}({\cal N}^{0}_{L}{\bf 1}+\Delta{\cal N}^{}_L\hat{\bf n}^{}_L\cdot\sig)\}\ .
\end{align}

In our model, the tunneling matrix has the form [see Eqs. (\ref{Vs})]
\begin{align}
&\widetilde{V}^{}_{LR}=V_0{\bf 1}+{\bf C}\cdot\sig\ ,\ \ \ {\rm with}\  \ {\bf C}=V_b\hat{\bf b}+iV_{\rm so}\hat{\bf b}_{\rm so}\ .
\end{align}
Using these notations, one finds
\begin{align}
\widetilde{V}^{}_{LR}\widetilde{V}^{\dagger}_{LR}&=V_0^2{\bf 1}+V_0({\bf C}+{\bf C}^*)\cdot\sig+|{\bf C}|^2+i[{\bf C}\times{\bf C}^*]\cdot\sig\nonumber\\
&=(V_0^2+V_b^2+V_{\rm so}^2){\bf 1}+2V_b\big(V_0\hat{\bf b}-V_{\rm so}[\hat{\bf b}_{\rm so}\times\hat{\bf b}]\big)\cdot\sig
=\mathcal{G}^{}_0{\bf 1}+(\mg^{}_1-\mg^{}_2)\cdot\sig
\ ,
\end{align}
where $\mathcal{G}^{}_0,~\mg^{}_1$ and $\mg^{}_2$ are defined in Eq. (\ref{calG}), and
\begin{align}
\widetilde{V}^{}_{LR}(\hat{\el}\cdot\sig)\widetilde{V}^{\dagger}_{LR}&=(V_0^2-|{\bf C}|^2)(\hat{\el}\cdot\sig)+(V_0[{\bf C}+{\bf C}^*]\cdot\hat{\el}+i[{\bf C}^*\times{\bf C}]\cdot\hat{\el}){\bf 1}\nonumber\\
&+iV_0[{\bf C}-{\bf C}^*]\times\hat{\el}\cdot\sig+[{\bf C}^*\cdot\hat{\el}{\bf C}+{\bf C}\cdot\hat{\el}{\bf C}^*]\cdot\sig 
=(\mg^{}_1+\mg^{}_2)\cdot\hat{\el}{\bf 1} +\mg^{\hat{\el}}_{3}\cdot\sig\ ,
\label{C4}
\end{align}
\end{widetext}
where $\mg^{\hat{\el}}_{3}$ is defined in Eq. (\ref{GG3}) and $\hat{\el}$ is an arbitrary unit vector.
Note that interchanging $L\leftrightarrow R$ reverses the sign of $\hat{\bf b}^{}_{\rm so}$, and therefore the signs of $\mg^{}_{2}$ and of the last term in $\mg^{\hat{\el}}_3$, leaving the other $\mathcal{G}$'s unchanged.
This change is also equivalent to interchanging  $\widetilde{V}^{}_{L:R}$ with $\widetilde{V}^\dagger_{RL}$, and therefore
\begin{align}
\widetilde{V}^\dagger_{LR}\widetilde{V}^{}_{LR}&=\mathcal{G}^{}_0{\bf 1}+(\mg^{}_1+\mg^{}_2)\cdot\sig\ ,\nonumber\\
\widetilde{V}^\dagger_{LR}(\hat{\el}\cdot\sig)\widetilde{V}^{}_{LR}&=(\mg^{}_1-\mg^{}_2)\cdot\hat{\el}{\bf 1} +\mg^{\hat{\el}}_{3'}\cdot\sig\ ,
\label{C5}
\end{align}
with
\begin{align}
\mg^{\hat{\el}}_{3'}&=(2V_0^2-\mathcal{G}^{}_0)\hat{\el}+2\Big(V_b^2(\hat{\bf b}\cdot\hat{\el})\hat{\bf b}+V_{\rm so}^2(\hat{\bf b}_{\rm so}\cdot\hat{\el})\hat{\bf b}^{}_{\rm so}\nonumber\\
&+V_0V_{\rm so}[\hat{\bf b}^{}_{\rm so}\times\hat{\el}]\Big)\ .
\label{3ps}
\end{align}
These identities are sufficient for calculating all the necessary traces.

The `longitudinal' component of the magnetization rate is
\begin{align}
\dot{M}^L_\parallel=\dot{\bf M}^{L}_{}\cdot\hat{\bf n}^{}_L&=
G^{}_{L,\parallel}
(\mu^{0}_{R}-\mu^{0}_{L})\nonumber\\
&
-G^{}_{0}U^{}_{L}+G^{}_{\times ,\parallel}U^{}_{R}\ ,
\label{Mz}
\end{align}
and the only additional coefficient is $G^{}_{\times,\parallel}$, given in Eq. (\ref{Gxpar}).
The `transverse' component is needed only when the left lead in unpolarized, namely when $\Delta{\cal N}^{}_L=0$, and the result is given in Eq. (\ref{Mx}).

The coefficients in Eq. (\ref{Gz}) are found to be
\begin{align}
G^{}_0&=4\pi J^2\big({\cal N}^0_L{\cal N}^0_R\mathcal{G}^{}_0+{\cal N}^0_L\Delta{\cal N}^{}_R(\mg_1+\mg^{}_2)\cdot\hat{\bf n}^{}_R\nonumber\\
&+{\cal N}^0_R\Delta{\cal N}^{}_L(\mg_1-\mg^{}_2)\cdot\hat{\bf n}^{}_L+\Delta{\cal N}^{}_L\Delta{\cal N}^{}_R\mathcal{G}^{}_4\big)\ ,\nonumber\\
G^{}_{L,\parallel}&=4\pi J^2[{\cal N}^0_L{\cal N}^0_R(\mg_1-\mg^{}_2)\cdot\hat{\bf n}^{}_L+{\cal N}^0_L\Delta{\cal N}^{}_R\mathcal{G}^{}_4\nonumber\\
&+{\cal N}^0_R\Delta{\cal N}^{}_L\mathcal{G}^{}_0+\Delta{\cal N}^{}_L\Delta{\cal N}^{}_R(\mg_1+\mg^{}_2)\cdot\hat{\bf n}^{}_R]\ ,\nonumber\\
G^{}_{R,\parallel}&=4\pi J^2[{\cal N}^0_L{\cal N}^0_R(\mg_1+\mg^{}_2)\cdot\hat{\bf n}^{}_R+{\cal N}^0_L\Delta{\cal N}^{}_R\mathcal{G}^{}_0 \nonumber\\
&+{\cal N}^0_R\Delta{\cal N}^{}_L\mathcal{G}^{}_4+\Delta{\cal N}^{}_L\Delta{\cal N}^{}_R(\mg_1-\mg^{}_2)\cdot\hat{\bf n}^{}_L]\ ,
\label{GGGG}
\end{align}
where [see Eqs. (\ref{Vs})]
\begin{align}
\mathcal{G}^{}_0&=V_0^2+V^2_{\rm so}+V_b^2\ ,\nonumber\\
\mg^{}_{1}&=2V_bV_0 \hat{\bf b}\ ,\nonumber\\
\mg^{}_{2}&=2V_bV_{\rm so}\big[\hat{\bf b}_{\rm so}\times\hat{\bf b}\big]\ ,
\label{calG}
\end{align}
and
\begin{align}
\mathcal{G}^{}_4&=(2V_0^2-\mathcal{G}^{}_0)\hat{\bf n}^{}_R\cdot\hat{\bf n}^{}_L+2\Big(V_b^2\hat{\bf b}\cdot\hat{\bf n}^{}_R\hat{\bf b}\cdot\hat{\bf n}^{}_L\nonumber\\
&+V_{\rm so}^2\hat{\bf b}_{\rm so}\cdot\hat{\bf n}^{}_R\hat{\bf b}^{}_{\rm so}\cdot\hat{\bf n}^{}_L-V_0V_{\rm so}[\hat{\bf b}^{}_{\rm so}\times\hat{\bf n}^{}_R]\cdot\hat{\bf n}^{}_L\Big)\ .
\label{G4}
\end{align}
The  coefficient $\mathcal{G}^{}_4$ is even under interchanging $L$ by $R$.
Since $\Delta{\cal N}^{}_L$ and $\Delta{\cal N}^{}_R$ are odd under changing the sign of the lead magnetizations, and $\mg^{}_1$ and $\mg^{}_2$ are odd under changing the sign of $B$,
$G^{}_0$ is even and the other two coefficients are odd under such a change.
Also, as interchanging $L$ with $R$ changes the sign of $\mg^{}_{2}$, we find $G^{}_{L,\parallel}\leftrightarrow G^{}_{R,\parallel}$ under $L\leftrightarrow R$, implying that particle number is conserved.

The other coefficients of the magnetization rates are
\begin{align}
G^{}_{\times,\parallel}
&=4\pi J^{2}\big({\cal N}^0_L{\cal N}^0_R\mathcal{G}^{}_4+{\cal N}^0_L\Delta{\cal N}^{}_R(\mg^{}_1-\mg^{}_2)\cdot\hat{\bf n}^{}_L \nonumber\\
&+{\cal N}^0_R\Delta{\cal N}^{}_L(\mg^{}_1+\mg^{}_2)\cdot\hat{\bf n}^{}_R+\Delta{\cal N}^{}_L\Delta{\cal N}^{}_R\mathcal{G}^{}_0\big)\ ,
\label{Gxpar}
\end{align}
and (when $\Delta{\cal N}^{}_L=0$)
\begin{align}
{\bf G}^{}_{L}&=2\pi J^2_{}{\cal N}^0_L{\rm Tr}\{\widetilde{V}^{}_{LR}\NN^{}_R\widetilde{V}^\dagger_{LR}\sig\}\nonumber\\
&=4\pi J^2_{}{\cal N}^0_L\big[{\cal N}^0_R(\mg^{}_1-\mg^{}_2)+\Delta{\cal N}^{}_R
\mg^{{\bf n}^{}_R}_{3}\big]\ ,\nonumber\\
{\bf G}^{}_{\times}&=2\pi J^2_{}{\cal N}^0_L{\rm Tr}\{\widetilde{V}^{}_{LR}\NN^{}_R\sig\cdot\hat{\bf n}^{}_R\widetilde{V}^\dagger_{LR}\sig\}\nonumber\\
&=4\pi J^2_{}{\cal N}^0_L\big[{\cal N}^0_R\mg^{\hat{\bf n}^{}_R}_{3}+\Delta{\cal N}^{}_R(\mg^{}_1-\mg^{}_2)\big]\ ,
\label{Mx}
\end{align}
where
\begin{align}
\mg^{\el}_3&=(2V_0^2-\mathcal{G}^{}_0)\hat{\el}+2\Big(V_b^2(\hat{\bf b}\cdot\hat{\el})\hat{\bf b}+V_{\rm so}^2(\hat{\bf b}_{\rm so}\cdot\hat{\el})\hat{\bf b}^{}_{\rm so}\nonumber\\
&-V_0V_{\rm so}[\hat{\bf b}^{}_{\rm so}\times\hat{\el}]\Big)\ ,
\label{GG3}
\end{align}
for an arbitrary unit vector ${\el}$.
Interestingly, the total magnetization is not conserved, and we find that the link injects polarized spins into the leads,
$\dot{\bf M}^{L}+\dot{\bf M}^{R}\ne 0$.



\begin{thebibliography}{99}
\bibitem{winkler}
R. Winkler,
{\it Spin-Orbit Coupling Effects in Two-Dimensional Electron and Hole Systems},
 (Springer-Verlag, Berlin, 2003).
\bibitem{manchon}
A. Manchon, H. C. Koo, J. Nitta, S. M. Frolov, and R. A. Duine,
{\it New Perspectives for Rashba Spin-Orbit Coupling},
Nat. Mater. {\bf 14}, 871 (2015).
\bibitem{Kohda}
M. Kohda and J. Nitta,
{\it Enhancement of spin-orbit interaction and the effect of interface diffusion in quaternary InGaAsP/InGaAs heterostructures},
Phys. Rev. B {\bf 81}, 115118 (2010).
\bibitem{rashba} E. I. Rashba,
{\it Properties of semiconductors with an extremum loop .1. Cyclotron and combinational resonance in a magnetic field perpendicular to the plane of the loop}, Fiz. Tverd. Tela (Leningrad) {\bf 2}, 1224 (1960) [Sov.
Phys. Solid State {\bf 2}, 1109 (1960)]; Y. A. Bychkov and E. I. Rashba, {\it Oscillatory effects and the magnetic susceptibility of carriers in inversion layers},
J. Phys. C {\bf 17}, 6039 (1984).
\bibitem{Nitta}
J. Nitta, T. Akazaki, H. Takayanagi, and T. Enoki,
{\it
Gate Control of Spin-Orbit Interaction in an Inverted In$_{0.53}$Ga$_{0.47}$As/In$_{0.52}$Al$_{0.48}$As Heterostructure},
Phys. Rev. Lett. {\bf 78}, 1335 (1997).
\bibitem{Sato}
Y. Sato, T. Kita, S. Gozu, and S. Yamada,
{\it Large spontaneous spin splitting in gate-controlled two-dimensional electron gases at normal In$_{0.75}$Ga$_{0.25}$As/In$_{0.75}$Al$_{0.25}$As
heterojunctions},
J. Appl. Phys. {\bf 89}, 8017 (2001).
\bibitem{Beukman}
A. J. A. Beukman, F. K. de Vries, J. van Veen, R. Skolasinski,  M.  Wimmer,  F.  Qu,  D. T.  de  Vries,  B. M.  Nguyen,
W. Yi, A. A. Kiselev, M. Sokolich, M. J. Manfra, F. Nichele,
C. M.  Marcus,  and  L. P.  Kouwenhoven,
{\it Spin-orbit interaction in a dual gated InAs/GaSb quantum well},
Phys. Rev. B {\bf 96}, 241401 (2017).
\bibitem{comDres}
Other types of SOI, e.g. the Dresselhaus SOI [G. Dresselhaus, {\it Spin-Orbit Coupling Effects in Zinc Blende Structures}, Phys. Rev. {\bf 100}, 580 (1955)], or those resulting from strains in the wire [M. S. Rudner and E. I. Rashba, {\it Spin relaxation due to deflection coupling in nanotube quantum dots}, Phys. Rev. B {\bf 81}, 125426 (2010); K. Flensberg and C. M. Marcus, {\it Bends in nanotubes allow electric spin control and coupling}, Phys. Rev. B {\bf 81}, 195418 (2010),
G. A. Steele, F. Pei, E. A. Laird, J. M. Jol, H. B. Meerwaldt, and  L. P. Kouwenhoven, {\it
Large spin-orbit coupling in carbon nanotubes},
Nature Comm. 2584,  (2013)], yield similar effects, with other directions of the effective SOI magnetic field  around which the spins rotate.
\bibitem{datta}
S. Datta and B. Das,
{\it Electronic analog of the electro-optic modulator},
Appl. Phys. Lett. {\bf 56}, 665 (1990).
\bibitem{DDhistory} e.g., S. Sugahara and M. Tanaka, {\it A spin metal-oxide-semiconductor field-effect transistor using
half-metallic-ferromagnet contacts for the source and drain}, Appl. Phys. Lett. {\bf 84}, 2307 (2004) and references therein;
A. Hirohata and K. Takanashi, {\it Future perspectives for spintronic devices}, J.  Phys. D: Appl. Phys. {\bf 47}, 193001  (2014) and references therein.
\bibitem{recent} A recent experimental realization was reported by
S. Ringer, M. Rosenauer, T. V\"{o}lkl, M. Kadur, F. Hopperdietzel, D. Weiss, and J. Eroms, {\it Spin field-effect transistor action via tunable polarization of the spin injection in a Co/MgO/graphene contact}, Appl. Phys. Lett. {\bf 113}, 132403 (2018).
\bibitem{zutic1} For a recent review, see I. \v{Z}uti\'{c}, A. Matos-Abiague, B. Scharf, H. Dery, and K. Belashchenko, {\it Proximitized materials}, Materials Today {\bf 22}, 85 (2019). See also V. Lopez and E. V. Anda, {\it Spin polarized current in a quantum dot connected to a soin-orbit interacting sea}, J. Phys. Chem. Sol. {\bf 128}, 188 (2019). 
\bibitem{flensberg}
M. H. Larsen, A. M. Lunde, and K. Flensberg, {\it Conductance of Rashba spin-split systems with ferromagnetic contacts}, Phys. Rev. B {\bf 66}, 033304 (2002).
\bibitem{bardarson}
J. H. Bardarson,
{\it A proof of the Kramers degeneracy of transmission eigenvalues from antisymmetry of the scattering matrix},
J. Phys. A: Math. Theor. {\bf 41}, 405203 (2008).
\bibitem{Aharony_2018}
A. Aharony, O. Entin-Wohlman, M. Jonson, and R. I. Shekhter, {\it Electric and magnetic gating of Rashba-active weak links}, Phys. Rev. B {\bf 97}, 220404(R) (2018).
\bibitem{Aharony2_2019}  O. Entin-Wohlman and A. Aharony, {\it Spin geometric-phases in hopping magnetoconductance and spin currents}, Phys. Rev. Research {\bf 1}, 033112 (2019).
\bibitem{Shekhter_2013} R. I. Shekhter, O. Entin-Wohlman, and A. Aharony, {\it Suspended Nanowires as Mechanically Controlled Rashba Spin Splitters}, Phys. Rev. Lett. {\bf 111}, 176602 (2013); see also Supplemental
Material at http://link.aps.org/supplemental/10.1103/PhysRevLett.
111.176602.
\bibitem{Shekhter_2014} R. I. Shekhter, O. Entin-Wohlman and A. Aharony, {\it Mechanically controlled spin-selective transport}, Phys. Rev. B {\bf 90}, 045401 (2014).
\bibitem{Aharony1_2019} A. Aharony, O. Entin-Wohlman, K. Sarkar, R. I. Shekhter and M. Jonson, {\it Effects of Different Lead Magnetizations on the Datta-Das Spin Field Effect Transistor}, J. Phys. Chem. C {\bf 123}, 11094 (2019). The present paper corrects two errors in this reference:
 (a) Following Ref. \onlinecite{Aharony_2018}, the particle and spin-current expressions there have wrong sign, and there was an erronous factor of 2 in the second line of Eq. (19); (b) the second-order correction term from the leads was not considered. It was also missing in Ref. \onlinecite{Shekhter_2014}, but that paper only considered the `longitudinal' spin currents, which are not affected by this term.
\bibitem{keldysh} O. Entin-Wohlman, unpublished.
 \bibitem{Meir}
A-P. Jauho, N. S. Wingreen, and Y. Meir,
{\it Time-dependent transport in interacting and noninteracting resonant-tunneling systems},
Phys. Rev. B {\bf 50}, 5528 (1994).

\bibitem{raikh}
T. V. Shahbazyan and M. E. Raikh, {\it Low-Field Anomaly in
20 Hopping Magnetoresistance Caused by Spin-Orbit Term
in the Energy Spectrum}, Phys. Rev. Lett. {\bf 73}, 1408 (1994).
\bibitem{AC}
Y. Aharonov and A. Casher,
 {\it Topological quantum Effects for Neutral Particles},
Phys. Rev. Lett. {\bf 53}, 319 (1984).

\bibitem{69} D. Grundler, {\it Large Rashba Splitting in InAs Quantum Wells due to Electron Wave Function Penetration into the Barrier Layers}, Phys. Rev. Lett. {\bf 84}, 6074 (2000).
\bibitem{70} M. Kohda and J. Nitta, {\it Enhancement of spin-orbit interaction and the effect of interface diffusion in quaternary InGaAsP/InGaAs heterostructures}, Phys. Rev. {\bf 81}, 115118 (2010).



\end{thebibliography}
\end{document}